\documentclass[a4paper]{JAC2003}
\usepackage{graphicx}
\usepackage{booktabs}
\usepackage{varwidth}
\usepackage{xcolor}

\begin{document}
\title{LUMINOSITY LEVELLING TECHNIQUES FOR THE LHC}

\author{B. Muratori\thanks{bruno.muratori@stfc.ac.uk}, STFC Daresbury Laboratory, ASTeC and Cockcroft Institute, UK \\
T. Pieloni\thanks{tatiana.pieloni@cern.ch}, CERN, Geneva, Switzerland}

\maketitle

\begin{abstract}
We present the possibilities for doing luminosity levelling at the LHC. We explore the merits and drawbacks of each option and briefly discuss
the operational implications. The simplest option is levelling with an offset between the two beams. Crab cavities may also be used for
levelling, as may a squeezing of the beam. There is also the possibility of using the crossing angle in order to do luminosity levelling. All
of these options are explored, for the LHC and other possible new projects, together with their benefits and drawbacks.
\end{abstract}

\section{Introduction}
\begin{figure}[htb]
\begin{center}
\includegraphics[width=80mm]{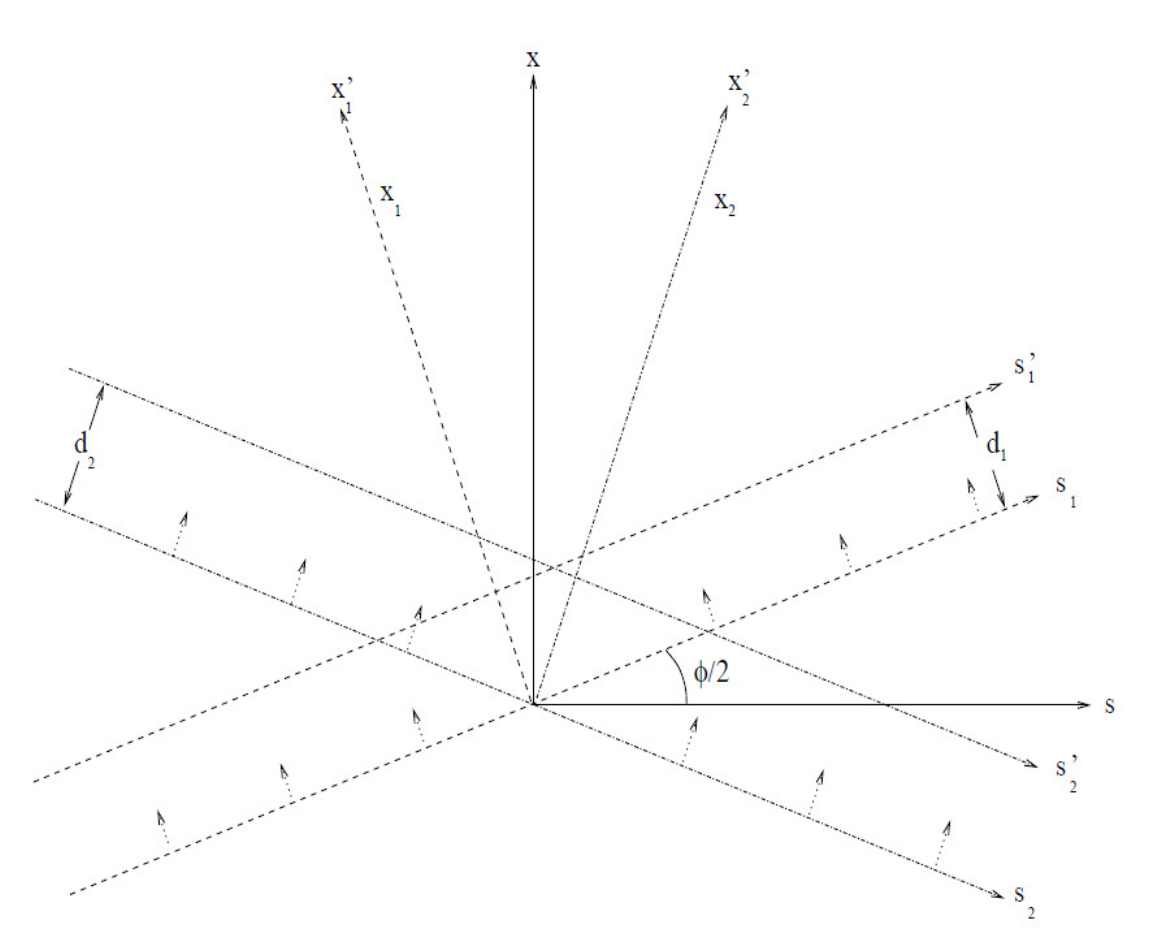}
\caption{The geometry of the interaction region, with the crossing angle and the offset of the beams.}
\label{cross}
\end{center}
\end{figure}

One of the main measures of a collider's performance is its luminosity. However, from the point of view of experiments, what is most important
is not the peak luminosity but, rather, the integrated luminosity. For the detection of events, it is also preferable that the luminosity remain
constant for as long as possible. Therefore, luminosity levelling can be introduced. This means that the natural decay of the luminosity is
pre-empted and the luminosity is spoilt initially with respect to the nominal.  Then, as the luminosity decays, it is spoilt less and less in
order that it remain constant for as long as possible. While doing this, it is still very much worthwhile to start with as high a luminosity
as possible, as this will translate in the luminosity being constant for a longer amount of time after levelling. To explain what is meant by
this, we consider the expression for the luminosity in the presence of both an offset and a crossing angle such that the crossing region is
illustrated by Fig. \ref{cross} (more details can be found in \cite{lumi,luminote}):
\[
{\cal L} = \frac{N_1N_2fN_b}{4\pi\sigma_x\sigma_y}W\mathrm{e}^{\frac{B^2}{A}}\frac{1}{\sqrt{1 + (\frac{\sigma_s}{\sigma_x}\tan\frac{\phi}{2})^2}},
\]
with
\[A = \frac{\sin^2\frac{\phi}{2}}{\sigma_x^2} + \frac{\cos^2\frac{\phi}{2}}{\sigma_s^2},
B = \frac{(d_2 - d_1)\sin\frac{\phi}{2}}{2\sigma_x^2},
\]
and
\[W = \mathrm{e}^{-\frac{1}{4\sigma_x^2}(d_2 - d_1)^2}.
\]
$N_1$ and $N_2$ are the number of protons per
bunch for Beams $1$ and $2$, respectively; $N_b$ is the number of colliding bunches per beam; $\sigma_x$, $\sigma_y$, and $\sigma_s$ are the
transverse and longitudinal bunch dimensions; $\phi$ is the crossing angle; and $d_1$ and $d_2$ are the offsets of Beams $1$ and $2$ with
respect to the nominal.

\begin{figure}[htb]
\begin{center}
\includegraphics[width=70mm]{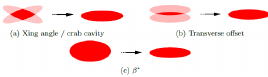}
\caption{A sketch of the luminous region for different levelling techniques.}
\label{ltypes}
\end{center}
\end{figure}

Various types of luminosity levelling have been suggested. These are explained below, together with an analysis of their merits and drawbacks,
as well as a discussion of how they satisfy the requirements, from the point of view of observation, operations, and the LHC experiments
\cite{papotti,jacobson,jacquet,giachino}. The main types of levelling are separation, crab cavities or crossing angle, and $\beta^*$ squeeze,
as illustrated in Fig. \ref{ltypes}.
\section{Levelling with offset}
The simplest form of levelling is achieved by introducing an offset between the two colliding beams. This is straightforward from an
operational point of view, and can be implemented easily and quickly if required. It also makes it possible to do levelling in all IPs
independently, as it is done with a local orbit bump, and it gives a smaller tune spread, therefore leading to smaller losses. The average number
of p-p collisions per bunch crossing, or pile-up, that experiments can handle is limited, as different events need to be distinguished. This
is particularly important longitudinally, where the vertex density is critical and levelling with offset allows for this to be kept constant.

\begin{figure}[htb]
\begin{center}
\includegraphics[width=70mm]{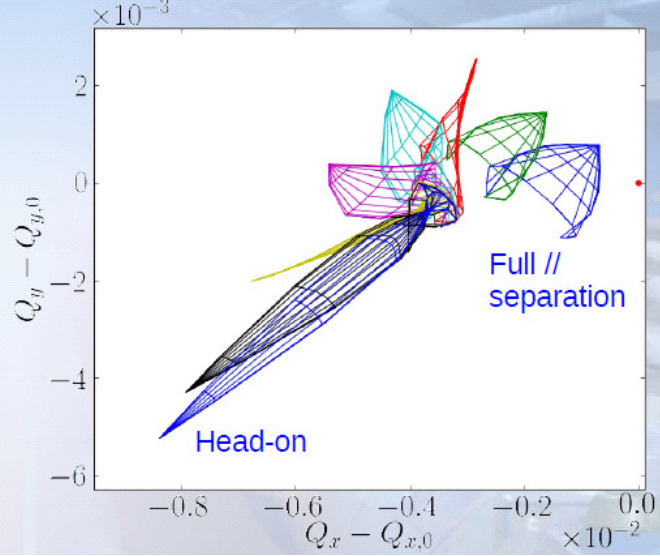}
\caption{The tune footprints for different separations.}
\label{tune}
\end{center}
\end{figure}

\begin{figure}[htb]
\begin{center}
\includegraphics[width=80mm]{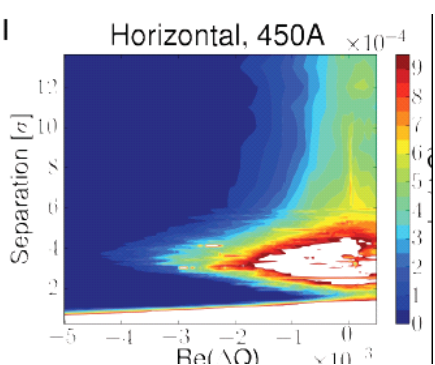}
\caption{The real tune shift (the extent of the stability region) as a function of the offset.}
\label{stability}
\end{center}
\end{figure}

\begin{figure}[htb]
\begin{center}
\includegraphics[width=80mm]{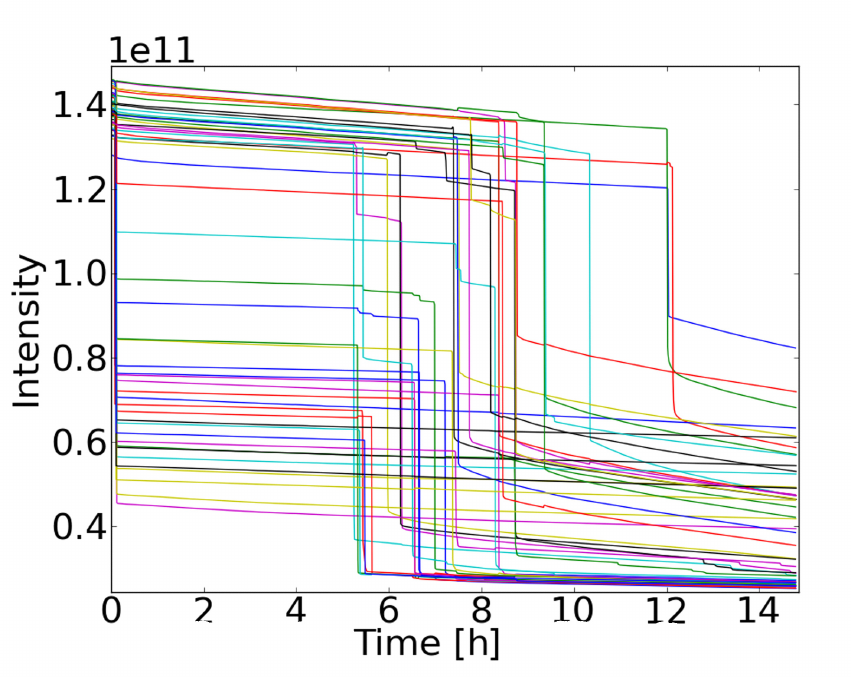}
\caption{The intensity of the IP8 bunches.}
\label{intensity}
\end{center}
\end{figure}

\begin{figure}[htb]
\begin{center}
\includegraphics[width=80mm]{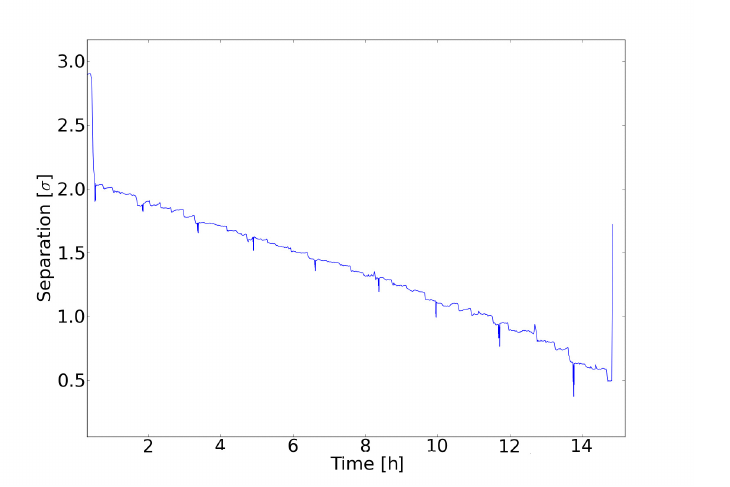}
\caption{Full separation in IP8.}
\label{separation}
\end{center}
\end{figure}

\begin{figure}[htb]
\begin{center}
\includegraphics[width=80mm]{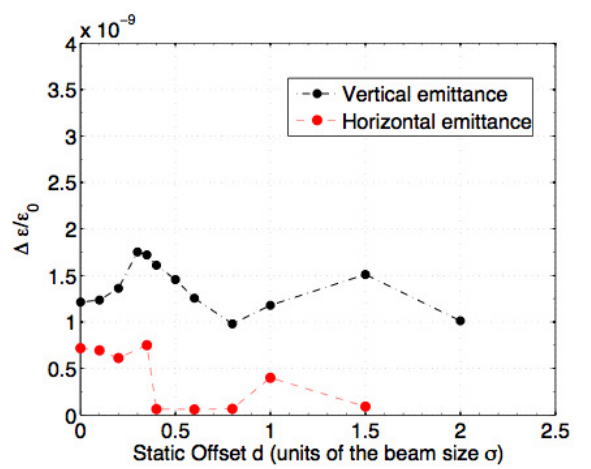}
\caption{Emittance as a function of beam separation.}
\label{emittance}
\end{center}
\end{figure}

Luminosity levelling with offset suffers from several drawbacks. The most obvious one is that a different separation leads to a different
beam--beam force being experienced. Therefore, the effect of one beam--beam encounter with a separation of a few orders of the r.m.s.\@ beam size can
change the tune spread appreciably, as shown in Fig. \ref{tune}. This leads to a decrease in the extent of the stability region as a
function of the offset, as shown in Fig. \ref{stability}. 
Note that the minimum is hardly visible and lies just below the big maximum shown in white. The position and amplitude of this minimum depends
on the collision schedule, the bunch intensities, emittances, octupole settings, and, in particular, the transverse offset at all the different
IPs and the bunches experiencing head-on collisions.  The fewer the head-on collisions, the smaller is the stability area shown in Fig.
\ref{stability}. In fact, as can be seen from the figure, there exists a critical separation at which the instability diagram is a minimum.
Therefore, it is believed that levelling with transverse offset leads to serious complications in ensuring the stability of all bunches
involved. Operationally \cite{MDnote}, it is believed that such effects have already been observed in IP8 where, during a few fills, bunches
that were colliding only in IP8 were lost and suffered substantial reductions in intensity, as shown in Fig. \ref{intensity}, together with
the full separation inferred from the measured luminosity in IP1 and IP8, given in Fig. \ref{separation}. Clearly, it is expected that this
effect will become even worse when bunches collide in more IPs and not all of them experience head-on collisions. Other drawbacks include the fact that the
tune shift keeps changing as the beams are brought into and out of collision, and that bunches become more sensitive to instabilities with
respect to head-on collisions. The mere fact of going into collision with a separation could in itself give rise to instabilities or maximize
their effect. Finally, there is a possible emittance growth resulting from the offsets used, as can be seen in Fig. \ref{emittance}.  

\section{Crab Cavity Levelling}
Crab cavities have been used successfully, in electron colliders, to increase the luminosity back to the nominal value in the presence of a
crossing angle.  In a similar way, they may be used to perform luminosity levelling by `spoiling' the luminosity initially, by artificially
`anti'-crabbing the beam, and subsequently by correcting for the natural exponential decrease in luminosity through the usual crabbing of
the beam. The main advantages of using crab cavities are that all IPs are independent, and that it is possible to go back and forth easily by
just changing the voltage of the cavities.

\begin{figure}[htb]
\begin{center}
\includegraphics[width=70mm]{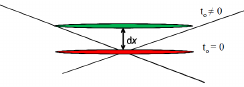}
\caption{The effect of crab cavity jitter at the IP, where $t_0$ is the time at which the bunch enters the cavity.}
\label{crab}
\end{center}
\end{figure}

Luminosity levelling with crab cavities suffers from several drawbacks. The most obvious one is that there is, so far, no experience with
applying crab cavities to proton bunches at all, and this would most likely lead to additional problems from an operational point of view.
The longitudinal vertex density changes with the levelled angle, giving rise to all the problems that were discussed above for offset
levelling. Further, the tunes change with the crossing angle, and additional noise could be introduced on the colliding beams, hence reducing the
reachable $\xi_{bb}$. Also, the jitter coming from the cavities needs to be dealt with. Differential phase jitter causes the two
bunches to have a height mismatch, which can significantly reduce luminosity or cause the bunches to miss. Phase jitter means that the entry
time of the centre of the bunch to is different for the cavities; hence d$x$ is different for the two beams, as may be seen in Fig. \ref{crab}. 

Phase jitter between the cavities causes the two beams to be displaced in the $x$-plane, which can reduce the luminosity of the collision or
even cause the bunches to miss each other completely. Further information about cavity phase jitter for the ILC can be found in \cite{burt}.
\section{$\beta^{*}$ Levelling}
\begin{figure}[htb]
\begin{center}
\includegraphics[width=80mm]{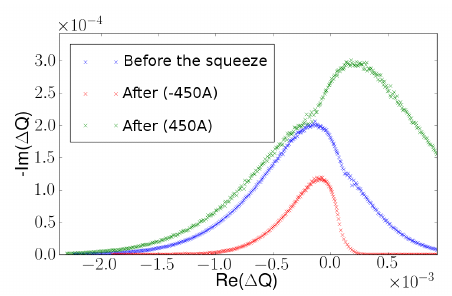}
\caption{The horizontal stability diagram for different octupole settings before and after the $\beta^*$ squeeze.}
\label{stab1}
\end{center}
\end{figure}

Another option for doing luminosity levelling is to start with a beam the cross-section of which is larger than the nominal and then gradually
squeeze it as the luminosity spontaneously reduces exponentially: this is known as $\beta^*$ levelling. Now, the stability of the beam
relies on impedance modes in the machine being Landau damped. This is done via a tune spread in the presence of beam--beam or other
non-linearities in the machine. In the absence of colliding beams, octupoles are used to ensure the required tune spread for damping and,
thereby, beam stability \cite{ruggiero}. Beam--beam effects may be safely ignored before the $\beta^*$ squeeze; however, during the squeeze,
the $\beta$ function grows dramatically in the region near the IP. This reduces the separation between the two beams even before they are
brought into collision. The stability region before and after the squeeze with two different octupole settings is shown in Fig. \ref{stab1}. 

\begin{figure}[htb]
\begin{center}
\includegraphics[width=80mm]{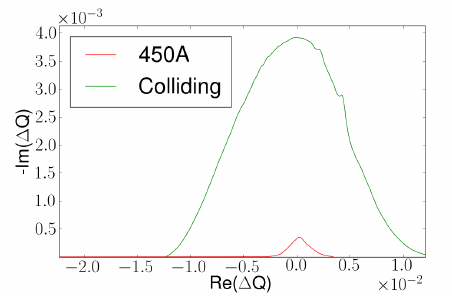}
\caption{The horizontal stability diagram with and without head-on collision.}
\label{stab2}
\end{center}
\end{figure}

Clearly, it is preferable to use the positive octupole polarity; however, the strength required means that there are some detrimental effects
associated with this, namely a reduction in dynamic aperture and feed-down effect. This is avoided if the squeeze is done when the beams are
already colliding head on. This ensures a much larger tune spread and hence Landau damping, giving a much larger stability diagram, as shown in
Fig. \ref{stab2}.

\begin{figure}[htb]
\begin{center}
\includegraphics[width=80mm]{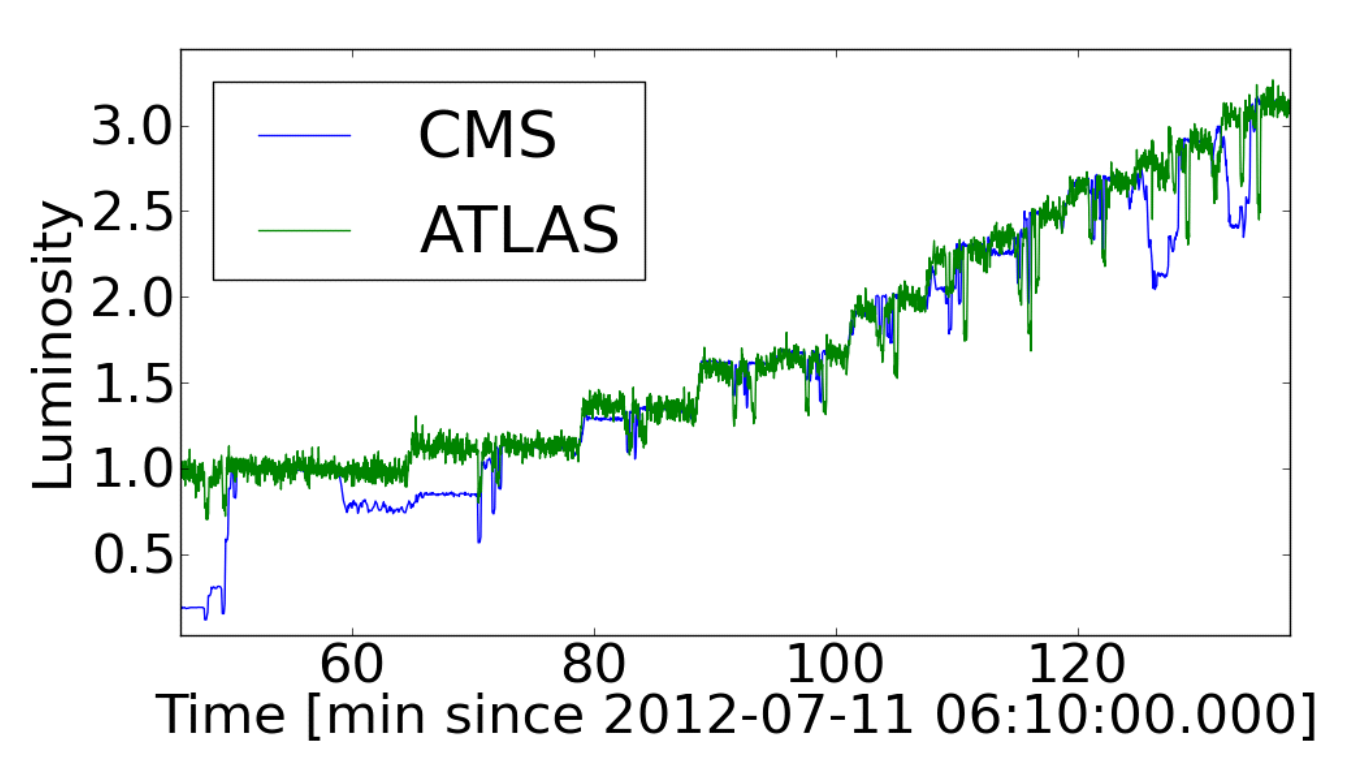}
\caption{The luminosity evolution during fill $2828$.}
\label{level1}
\end{center}
\end{figure}

\begin{figure}[htb]
\begin{center}
\includegraphics[width=80mm]{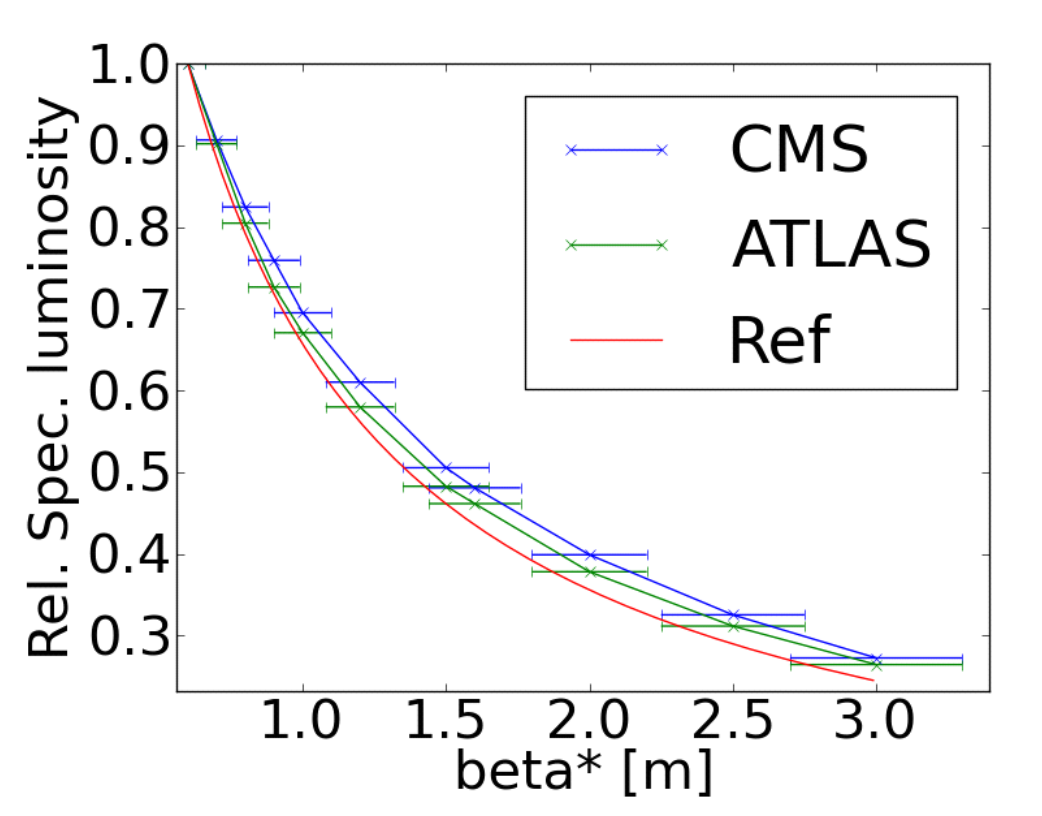}
\caption{The measured luminosity reduction factor at CMS and ATLAS, compared to expectations.}
\label{level2}
\end{center}
\end{figure}

The principle of $\beta^*$ levelling is illustrated in Fig. \ref{level1}, for an experiment done in 2012 \cite{MDnote} where the beam was
slowly squeezed, as a function of time, and the luminosity increased. Physically, there is no difference between this and keeping the
luminosity constant, as it naturally degrades. All parameters, such as beam size, tunes, and orbit, should be monitored while the luminosity is
being squeezed. The main advantages are as follows: there is a constant longitudinal vertex density for the experiments; the tunes do not change and
are constant over the fill; and it is more stable with the largest area of Landau damping.
As the tune spread from head-on beam--beam does not depend on $\beta^*$, leveling with it would allow a constant stability diagram to be maintained
during the procedure, as opposed to what happens when the levelling is done with just offset. Figure \ref{level2} shows a comparison of the
measured luminosity reduction factors when doing $\beta^*$ levelling for the experiment performed at the LHC in 2012 \cite{MDnote}, both at
CMS and ATLAS, as well as the expected reduction. 
The principal drawbacks are due to the orbit. This has to be kept constant during the squeeze, as the beams must be kept in collision, which
means that a feed-forward on it is required for robustness from an operational point of view and could require several changes from normal
operations.

\section{Other Levelling Possibilities}
Several other possibilities exist, and these are listed briefly below, Some of them are very new and have not been fully evaluated yet, while
others still require experimental verification and further studies of their viability.
\begin{itemize}
\item \emph{Longitudinal cogging}:  This means introducing a time delay of the order of a couple of RF periods longitudinally, thereby ensuring that there
is only a partial overlap of the two colliding bunches at the IP. So far, only $1$ or $5$ RF periods have been implemented experimentally. It
appears to be a relatively easy option to implement; however, it means that levelling is done at all IPs simultaneously and this is very
restrictive for the experiments. Longitudinal cogging also moves the luminous region longitudinally.
\item \emph{The large crossing (Piwinski) angle option}: This is where the levelling is actually done with a variation of the crossing angle. This
option also varies the length of the luminous region according to the crossing angle.
\item \emph{The flat beam option}: This has been proposed recently \cite{burov} and involves doing the levelling in one plane only, the same as the
crossing angle plane. This means that the tune shift in the other plane can be kept constant and the collimators do not have to move much,
which could otherwise lead to safety issues.
\end{itemize}
\section{Discussion}
Various scenarios for luminosity levelling, all valid working options, have been presented and their merits and drawbacks have been discussed. The
easiest to implement is the offset option; however, this could lead to instabilities. Crab cavities introduce an additional complexity, which
could turn out to be very non-trivial. There is also no experience of crab cavities and proton bunches, and the cavities could also introduce a
substantial jitter. Several other options, such as levelling with a crossing angle, flat beam options, or longitudinal cogging, have also been
discussed. However, it appears that $\beta^*$ levelling, possibly together with some offset as well, is the most promising option, as it appears
to satisfy most of the requirements, both from the experiments' point of view and operationally. The main problem with $\beta^*$ levelling
is that the orbit has to be kept constant as the levelling is being done, and this may be rather complex from an operational point of view.

Ultimately, the most important thing that will determine exactly how the luminosity shall be levelled, both at the LHC and in possible new
projects \cite{pieloni}, is the operational simplicity with which the method can be implemented, together with the experimental requirements
and constraints.
\section{Acknowledgements}
The authors would like to thank the Operation group for all the support during the testing of the $\beta^*$ levelling experiments, in particular W. Herr, J. Wenninger, S. Redaelli, and M. Lamont.


\begin{thebibliography}{99} 
\bibitem{lumi} W. Herr and B. Muratori, ``Concept of Luminosity,'' in Proc. CERN Accelerator School, Intermediate Level Course, Zeuthen, Germany, 2003, CERN-2006-002 (2006).
\bibitem{luminote} B. Muratori, ``Luminosity and Luminous Region Calculations for the LHC,'' LHC Project Note 301 (2002).
\bibitem{papotti} G. Papotti, ``Observations of Beam--Beam Effects in the LHC,'' these proceedings.
\bibitem{jacobson} R. Jacobsson, ``Needs and Requirements from the LHC Physics Experiments,'' these proceedings.
\bibitem{jacquet} D. Jacquet, ``Implementation and Experience with Luminosity Levelling with Offset Beams,'' these proceedings.
\bibitem{giachino} R. Giachino, ``Diagnostics Needs for Beam--Beam Studies and Optimization,'' these proceedings.
\bibitem{MDnote} X. Buffat et al., ``Results of $\beta^*$ Luminosity Leveling MD,'' CERN-ATS-Note-2012-071 MD (2012).
\bibitem{burt} A. Dexter et al., ``ILC Crab Cavity Phase Control System Development and Synchronisation Testing in a Vertical Cryostat Facility,'' EUROTeV-Report-2008-073 (2008).
\bibitem{ruggiero} J.P. Koutchouk and F. Ruggiero, ``A Summary on Landau Octupoles for the LHC,'' LHC Project Note 163 (1998).
\bibitem{burov} A. Burov, ``Circular Modes,'' these proceedings.
\bibitem{pieloni} T. Pieloni, ``Beam--Beam Studies in the LHC and New Projects,'' these proceedings.
\end{thebibliography}
\end{document}